\documentclass{emulateapj}

\slugcomment{Accepted to ApJ}

\shorttitle{Destruction of inner planets}
\shortauthors{Mustill, Davies, \& Johansen}

\begin{document}

\title{The destruction of inner planetary systems during\\high-eccentricity migration of gas giants}

\author{Alexander J. Mustill, Melvyn B. Davies, and Anders Johansen}
\affil{Lund Observatory, Department of Astronomy \& Theoretical Physics, Lund University, Box 43, SE-221 00 Lund, Sweden}
\email{alex@astro.lu.se}

\begin{abstract}
Hot Jupiters are giant planets on orbits a few hundredths of an AU. They do not share their system with low-mass close-in planets, despite these latter being exceedingly common. Two migration channels for hot Jupiters have been proposed: through a protoplanetary gas disc or by tidal circularisation of highly-eccentric planets. We show that highly-eccentric giant planets that will become hot Jupiters clear out any low-mass inner planets in the system, explaining the observed lack of such companions to hot Jupiters. A less common outcome of the interaction is that the giant planet is ejected by the inner planets. Furthermore, the interaction can implant giant planets on moderately-high eccentricities at semimajor axes $<1$ AU, a region otherwise hard to populate. Our work supports the hypothesis that most hot Jupiters reached their current orbits following a phase of high eccentricity, possibly excited by other planetary or stellar companions.
\end{abstract}


\section{Introduction}

Hot Jupiters were among the first exoplanets to be discovered \citep{MayorQueloz95}. However, their origin is still not understood, and models for their migration history fall into two categories: ``Type~II'' migration at early times through the protoplanetary gas disc \citep{Lin+96,Ward97}; and migration at late times as planets' eccentricities are excited by gravitational scattering in packed multi-planet systems \citep{RF96,Chatterjee+08}, or by secular perturbations from more distant planetary or binary companions \citep{WuMurray03,WuLithwick11,BeaugeNesvorny12,Petrovich14,Petrovich15}. High-eccentricity migration may better explain the observed misalignments between stellar spin and planetary orbits \citep{Triaud+10,Winn+10,WuLithwick11,BeaugeNesvorny12,Storch+14} as well as the innermost semi-major axes of the bulk of the hot Jupiter population \citep{FR06,PlavchanBilinski13,VR14}. A requirement of this channel is that hot Jupiters have (or had in the past) planetary or stellar companions on wide orbits, and indeed recent studies estimate that around 70\% of hot Jupiters have companion giant planets or stars on wide orbits \citep{Knutson+14,Ngo+15}.

On the other hand, hot Jupiters are not found to have low-mass, close-in companions. No such companions have yet been found by radial-velocity surveys, while survey results from the \emph{Kepler} spacecraft found no evidence of additional transiting companions or transit timing variations in hot Jupiter systems \citep{Steffen+12}; this deficit was statistically significant compared to multiplicities of warm Jupiter and hot Neptune systems. Nor have ground-based searches for companions that may cause strong transit timing variations proved fruitful \citep[e.g.,][]{Hoyer+12,Maciejewski+13}, despite these being sensitive to Earth-mass companions in mean motion resonance with a hot Jupiter. However, low-mass planets on close orbits are extremely common around stars that do not host hot Jupiters: results from {\it Kepler} transit photometry show that 52\% of stars have at least one planet with $P<85$ days and $R_\mathrm{pl}>0.8R_\oplus$ \citep{Fressin+13}; while radial-velocity surveys similarly show that 23\% of stars host at least one planet with $P<50$ days and $m_\mathrm{pl}>3M_\oplus$ \citep{Howard+10}. Furthermore, such planets often occur in multiple systems: the statistics of \textit{Kepler} candidate multiplicities requires a significant contribution from multi-planet systems \citep{Lissauer+11,FangMargot12,Fressin+13}. In many systems, then, migrating giant planets that will become hot Jupiters will interact with formed or forming systems of low-mass planets.

The lack of close companions to hot Jupiters can help to distinguish the different migration modes \citep{Steffen+12}. Simulations show that a giant planet migrating through an inner gas disc to become a hot Jupiter does not necessarily suppress planet formation in the inner disc \citep{MandellSigurdsson03,FoggNelson05,FoggNelson07a,FoggNelson07b,FoggNelson09,Mandell+07}, while embryos migrating after the giant form a resonant chain behind it and may accrete into a planet of detectable size \citep{Ketchum+11,Ogihara+13,Ogihara+14}.

In contrast, we show in this paper that during high-eccentricity migration, the giant planet almost always destroys all low-mass planets on orbits of a few tenths of an AU. Previous studies have shown that scattering among multiple giant planets can clear out material in the terrestrial planet region around 1\ AU, through direct scattering \citep{VerasArmitage05,VerasArmitage06} or secular resonance sweeping \citep{Matsumura+13}, and that it can suppress terrestrial planet formation in this region \citep{Raymond+11,Raymond+12}. We choose to focus our attention on very close-in systems more relevant for comparison to \textit{Kepler} observation ($\sim0.1$\ AU), which may have significant mass in inner planets (up to $\sim40\mathrm{\,M}_\oplus$ in total). We further consider the general case of a highly-eccentric giant planet, which may represent the outcome of scattering but which may also arise through other eccentricity excitation mechanisms such as Kozai perturbations or other secular effects.

In Section~2 of this paper we briefly review the population of planet candidates revealed by the \textit{Kepler} spacecraft. In Section~3 we describe the numerical approach we take to study the interaction of eccentric giant planets with close-in inner planets. In Section~4 we present the results of our numerical integrations. We discuss our findings in Section~5, and conclude in Section~6.

\section{Planetary multiplicities}

We show the multiplicities of the population of \textit{Kepler} planet candidates by taking the catalogue of \textit{Kepler} Objects of Interest (KOIs) from the Q1--Q16 data release at the NASA Exoplanet Archive (NEA) http://exoplanetarchive.ipac.caltech.edu/ (release of 2014-12-18; accessed 2015-01-08). This provided a list of 7348 planet candidates. KOIs may be genuine planets or false positives, with false positive probabilities up to 1 in 3 in some regions of parameter space \citep{Santerne+12,Coughlin+14}. Moreover, parameters for some planets in the NEA are unphysical.  We therefore performed several cuts on this list to attempt to remove false positives and poorly-characterised candidates:

\textit{No FPs:} Removal of any candidate classed as a false positive in the NASA Exoplanet Archive (in either of the columns ``disposition using \textit{Kepler} data'' or ``Exoplanet Archive disposition''). \textit{5739 candidates.}

\textit{L+11:} Following \cite{Lissauer+11}, we consider only planets with $SNR>16$, $P<240$ days and $R<22.4R_\oplus$, thus ensuring completeness and removing candidates with unphysically large radii. \textit{3678 candidates.}

\textit{L+11 \& no FPs:} Applies the cuts from \cite{Lissauer+11}, and also removes any false positives identified in the NEA Q1--Q16 data. \textit{3228 candidates.}

\textit{NEA good:} Removes NEA-identified false positives, and furthermore only includes planets that are listed as ``confirmed'' or ``candidate'' in at least one of the disposition columns, ensuring that the planets, if not confirmed, have passed some vetting to ensure a low probability of a false positive. \textit{2052 candidates.}

\begin{figure*}
  \includegraphics[width=.95\textwidth]{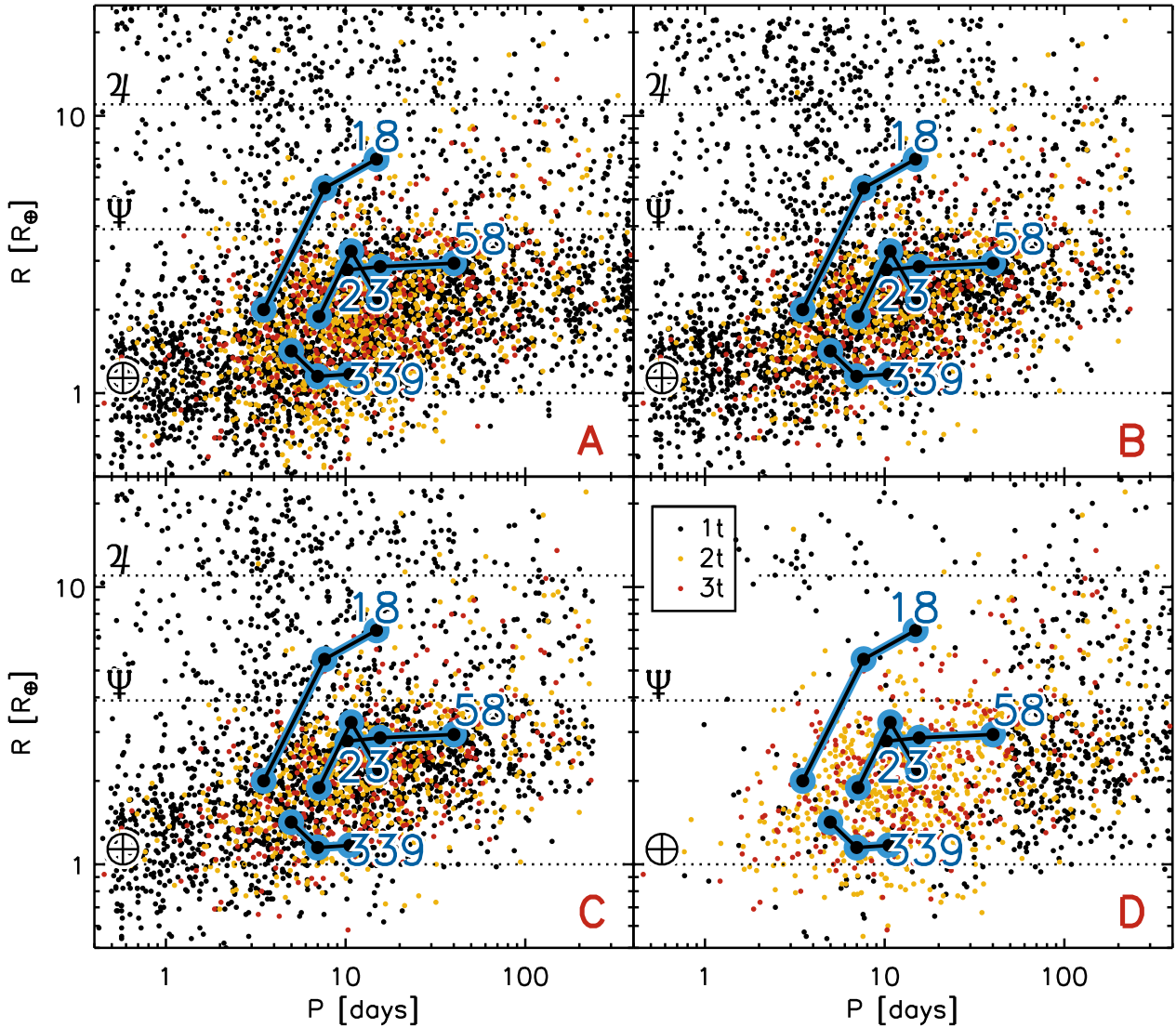}
  \vspace{.1in}
  \caption{\textbf{Periods and radii of candidate \emph{Kepler} single-, double- and triple-transit systems, with various selection criteria. A:} \textit{No FPs}---Known false positives removed; \textbf{B:} \textit{L+11}---Cuts on radius, SNR and period applied following \cite{Lissauer+11}; \textbf{C:} \textit{L+11 \& no FPs}---\cite{Lissauer+11} cuts applied and false positives removed; \textbf{D:} \textit{NEA good}---Only retains those candidates confirmed or with favourable vetting. See text for further details. Hot Jupiters are alone in systems, in contrast to both smaller planets and giant planets at longer orbital periods. We adopt the \textit{L+11 \& no FPs} selection as the most reliable. We mark the triple-candidate systems selected for our N-body integrations in blue, with their \emph{Kepler} numbers shown.}
  \label{fig:kois}
\end{figure*}

Plots of planet radius versus period for these four samples are shown in Fig~\ref{fig:kois}, where the contrast between solitary hot Jupiters and sociable low-mass planets is apparent. Although the numbers of single versus multiple systems vary \citep[in particular, the \textit{NEA good} sample has a great many multiples, as it is heavily influenced by the validation of numerous multiple-candidate systems by][]{Rowe+14}, for our purposes the key observation is that hot Jupiters at the top left of the plot are single. We do see some candidate hot Jupiters with companions, but these detections are not robust. For example, with the \textit{L+11} and \textit{L+11 \& no FPs} cuts, we find KOI-199.01 and KOI-199.02. The latter component is marked as a background eclipsing binary in the Q1--Q6 data from the NEA. In the \textit{NEA good} sample, we find KOI-338, confirmed by \cite{Rowe+14} as Kepler-141. This object has an unphysically large stellar radius in the NEA, measuring $19R_\odot$, larger than each of its planets' orbital radii. \cite{Rowe+14} assign a radius of $0.8R_\odot$, reducing the radii of the planet candidates proportionately. Hence, we do not consider either of these potential exceptions to our assertion that hot Jupiters are single to be reliable.

We adopt the \textit{L+11 \& no FPs} sample as the most reliable. This has 3228 planet candidates, forming 2136 single-planet systems, 282 doubles, 109 triples, 31 quadruples, 13 quintuples and 2 sextuples. We restrict attention to the triples and lower multiplicites as they offer better statistics.

\section{Numerical Method}

\subsection{N-body Model}

We conduct an extensive ensemble of N-body integrations with the Mercury package \citep{Chambers99}. We consider a highly-eccentric giant planet interacting with systems of three low-mass planets at $\sim0.1$ AU, chosen from among \textit{Kepler} triple-candidate systems, assuming that the three transiting planets are the only ones present in the inner system. Our systems are representative of the range of planet sizes of the multi-planet \textit{Kepler} systems (Fig~\ref{fig:kois}). Integrations are run for 1\,Myr.

We adopt the Bulirsch-Stoer algorithm with an error tolerance of $10^{-12}$. Within the 1\,Myr integration duration, energy conservation is generally good; we reject a small number of runs with $\Delta E/E>10^{-3}$. Collisions between bodies are treated as perfect mergers, and we consider a planet ejected from the system if it reaches a distance of 10\,000 AU from the star. The code does not incorporate general-relativistic corrections, but this is unimportant as the dynamics is dominated by scattering.

For our main integration runs, we take three-planet systems from the \textit{Kepler} triples and add to the system a highly-eccentric giant planet with a small pericentre. Our exemplar \textit{Kepler} systems are Kepler-18, Kepler-23, Kepler-58 and Kepler-339. Kepler-18, -23 and -339 all have planets with orbits from $\sim0.05$ to $\sim0.12$ AU, and span the range of planetary radii of the \textit{Kepler} multiple systems. Kepler-58 has planets on somewhat wider orbits, 0.09--0.23 AU.
These systems are marked in the space of \textit{Kepler} candidates in Fig~\ref{fig:kois}.

\begin{deluxetable}{lccc}
  \tabletypesize{\scriptsize}
  \tablecaption{Stellar parameters used in our integrations\label{tab:stars}}
  \tablewidth{0pt}
  \tablehead{
    \colhead{Name} & \colhead{$M_\star/M_\odot$} & \colhead{$R_\star/R_\odot$} & \colhead{Reference}
    }
  \startdata
  Kepler-18 & 0.972 & 1.108 & 1\\
  Kepler-23 & 1.11 & 1.52 & 2\\
  Kepler-58 & 0.95 & 1.03 & 3\\
  Kepler-339 & 0.902 & 0.802 & 4
  \enddata
  \tablerefs{(1) \cite{Cochran+11}; (2) \cite{Ford+12}; (3) \cite{Steffen+13}; (4) \cite{Rowe+14}}
\end{deluxetable}

\begin{deluxetable}{lcccc}
  \tabletypesize{\scriptsize}
  \tablecaption{Planetary parameters used in our integrations\label{tab:planets}}
  \tablewidth{0pt}
  \tablehead{
    \colhead{Name} & \colhead{$a/\mathrm{AU}$} & \colhead{$M_{pl}/M_\oplus$} & \colhead{$R_{pl}/R_\oplus$} & \colhead{Reference}\\
  }
  \startdata
  Kepler-18 b & 0.0477 & 6.9\tablenotemark{a} & 2.00 & 1\\
  Kepler-18 c & 0.0752 & 17.3 & 5.49 & 1\\
  Kepler-18 d & 0.1172 & 16.4 & 6.98 & 1\\
  \hline
  Kepler-23 b & 0.0749 & 4.86\tablenotemark{a} & 1.89 & 2\\
  Kepler-23 c & 0.0987 & 8.05\tablenotemark{a} & 3.25 & 2\\
  Kepler-23 d & 0.125 & 5.60\tablenotemark{a} & 2.20 & 2\\
  \hline
  Kepler-58 b & 0.0909 & 18.0 & 2.78 & 3\\
  Kepler-58 c & 0.1204 & 17.5 & 2.86 & 3\\
  Kepler-58 d & 0.2262 & 7.33\tablenotemark{a} & 2.94 & 4\\
  \hline
  Kepler-339 b & 0.0551 & 3.76\tablenotemark{a} & 1.42 & 4\\
  Kepler-339 c & 0.0691 & 1.74\tablenotemark{a} & 1.15 & 4\\
  Kepler-339 d & 0.0910 & 1.86\tablenotemark{a} & 1.17 & 4
  \enddata
  \tablerefs{(1) \cite{Cochran+11}; (2) \cite{Ford+12}; (3) \cite{WuLithwick13}; (4) \cite{Rowe+14}}
  \tablenotetext{a}{Mass is not measured directly: estimated from a mass--radius or density--radius relation \citep{WM14}.}
\end{deluxetable}

The \textit{Kepler} photometry allows a direct determination only of planet radii, but masses are more significant dynamically. Where available, we have taken masses determined by transit timing variations or radial velocities. Where these are unavailable, we have estimated masses based on a mass--radius or density--radius relation \citep{WM14}.
System parameters used for the simulations are given in Tables~\ref{tab:stars} and~\ref{tab:planets}.

For each of these systems, we conducted integration suites with different properties of the giant planet. For all four systems, we conducted integrations with the giant planet's initial semi-major axis of 10\,AU, while for Kepler-18 and -339 we also conducted integrations starting at 1.25\,AU. Within each combination of system and semi-major axis, we conducted 21 sets of 256 integrations, one set for each pericentre value $q$ from $0.01$ AU to $0.20$  in steps of $0.01$ AU, and a final set at $0.25$ AU (see Fig~\ref{fig:bars}). Within each set, half of the giants were on prograde and half on retrograde orbits; within each subsample, the orientation of the orbit was isotropic in the respective hemisphere. The giant was always released from apocentre. The giant's mass and radius were set to Jupiter's values. Our set-up assumes that during the initial excitation of the giant planet's eccentricity, there is no effect on the inner system,  a reasonable assumption \citep[for example, a tightly-packed system of planets protects itself against the Kozai effect,][]{Innanen+97}.

For the inner systems, we assigned the planets initially circular orbits with inclinations of up to $5^\circ$ from the reference plane, giving a maximum mutual inclination of $10^\circ$ with the distribution peaking at around $3.5^\circ$ \citep{Johansen+12}. We conducted an additional integration suite for the Kepler-18 system starting from a highly-coplanar configuration of inner planets (inclinations up to $0.001^\circ$), finding little impact on the outcome. The initial orbital phases of the inner planets were randomised. We conducted some integrations without giant planets to verify that our three-planet systems do not destabilise themselves on relevant timescales. No unstable systems were found over 1\,Myr (128 runs for each \textit{Kepler} triple studied).

We also conduct some ancillary integrations to test the effects of the relative orbital energies of the inner planets and the giant planet on the probability of ejecting the giant. All of these integrations were performed with a semimajor axis of 10 AU and a pericentre of 0.02\,AU for the giant. We tested two hot Jupiter systems (51 Pegasi, \citealt{MayorQueloz95,Butler+06}; and HAT-P-7, \citealt{Pal+08}) and one high-multiplicity system discovered by radial velocity \citep[$\tau$ Ceti;][]{Tuomi+13}. We also conducted additional integrations for Kepler-18 with the giant's mass set to $0.1$, $0.3$ and $3M_J$ at 10 AU, and with the giant's mass set to $1M_J$ at 5 and $2.5$ AU. The ejection fractions from these integrations are used in the discussion of the effects of orbital energy on ejection probability (see \S4), but their statistics are not otherwise discussed.

\subsection{Tidal Model}

Although we do not incorporate tidal forces into our N-body integrations, we post-process the planets surviving at the end of the 1\,Myr N-body integration to follow their orbital evolution under tidal forces. To model the tidal evolution of the planets after the interaction between the giant and the inner planets has concluded, we use the simple ``constant $Q$'' model in the form given in \cite{Dobbs-Dixon+04}. We include only the planetary tide, which is the most important for the planets' eccentricity decay until the host star leaves the Main Sequence \citep{Villaver+14}. We adopt values for the planets' tidal quality factors of $Q_\mathrm{pl}^\prime=10^6$ for the giants; $Q_\mathrm{pl}^\prime=10^5$ for the ``Neptunes'' Kepler-18c, d and their merger products; and $Q_\mathrm{pl}^\prime=10^2$ for the super-Earth Kepler-18b. These values are at the high end of those estimated for Solar System giants \citep{GoldreichSoter66} but comparable to estimates for exoplanets \citep{Jackson+08}. We tidally evolve our systems for 10\,Gyr. We note that observed systems may have had less time to tidally evolve, and a shorter evolution with a proportionally smaller $Q$ will give the same outcome.

\newpage

\section{Results}

\begin{figure*}
  \includegraphics[width=.95\textwidth]{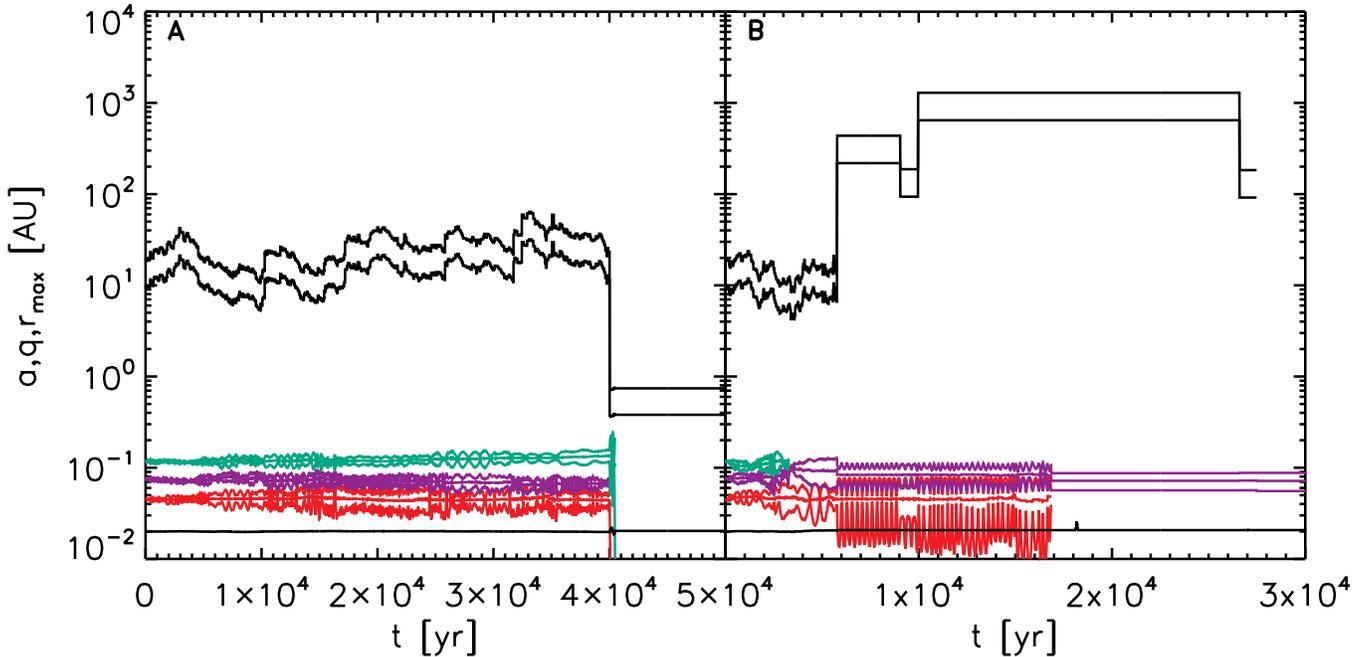}
  \caption{\textbf{Example orbital evolution from our integrations.} Semimajor axis ($a$), pericentre ($q$) and apocentre ($r_\mathrm{max}$) are shown for the giant and the three inner planets in two runs from the Kepler-18 integrations. The giant starts with a pericentre of 0.02 AU, meaning its orbit is overlapping those of all the inner planets. \textbf{A:} the eccentricities of the inner planets are stirred up, while the giant planet's semimajor axis fluctuates as a result of energy exchange with the inner planets. Just after 40 kyr, the giant planet and planet c collide; shortly after that, the remaining inner planets are driven into the star. The outcome is the giant planet surviving alone on a more tightly-bound orbit, with little overall change in its pericentre. \textbf{B:} the inner planets' orbits begin intersecting sooner, and c and d collide just after 3 kyr. Around 6 kyr, the giant is kicked onto a very wide orbit, only completing a few more orbits before acquiring a hyperbolic orbit at around 27 kyr. Meanwhile the c--d merger product swallows planet b, leaving a single inner planet in the system after ejection of the eccentric giant.}
  \label{fig:examples}
\end{figure*}

Our N-body simulations show that, in most cases, the systems resolve to one of two outcomes on time-scales much shorter than the integration duration: either all of the inner planets are destroyed (usually by collision with the star), leaving a single eccentric giant; or the giant is ejected by the inner planets, leaving 1--3 inner planets in the system, all of low mass. Examples of orbital evolution leading to these outcomes are shown in Fig~\ref{fig:examples}.

\begin{figure*}
  \includegraphics[width=\textwidth]{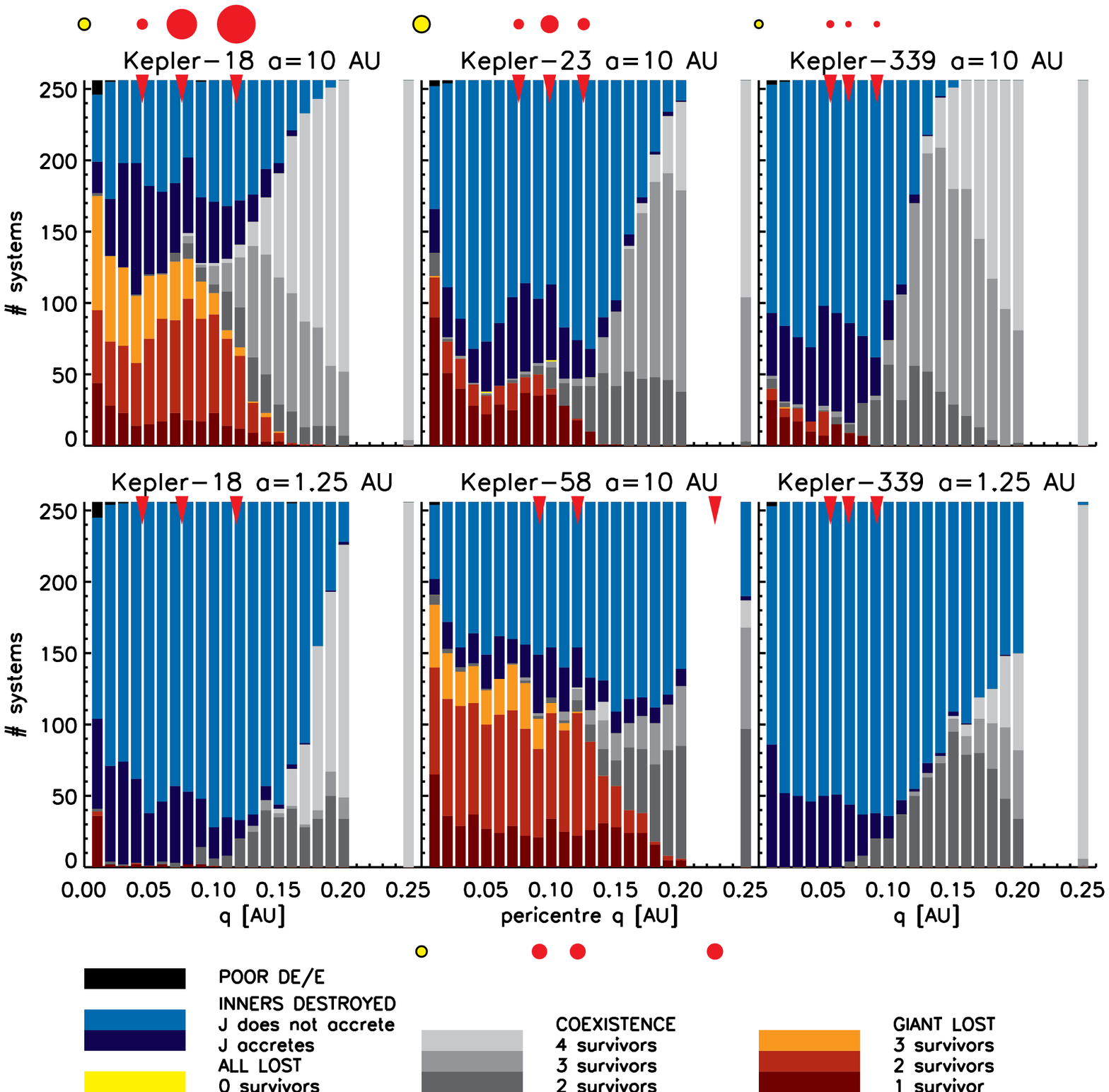}
  \caption{\textbf{Outcomes of our simulations, after 1 Myr of orbital evolution, shown as stacked bars as a function of the giant planet's initial pericentre $q$.} Blue shows systems where the giant destroys the inner planets (possibly by accreting one or more of them), red/orange shows systems where the giant is lost (usually by ejection, although collisions with the star can occur at $q=0.01$\ AU) and greys show systems where the giant and at least one inner planet coexist after 1\,Myr. In a small number of systems, all planets are lost (yellow), or energy conservation is poor (black). Red triangles mark the initial semimajor axes of the inner planets. Yellow and red circles show the systems to scale (planet radii inflated by a factor of 50). When the giant's pericentre comes inside the semi-major axis of the outermost inner planet, most integrations result in ejection of the giant or destruction of the inner planets. Destruction of the inner planets is favoured unless the inner planets are very massive (as in Kepler-18 and -58) and the giant is weakly bound ($a=10$\,au).}
  \label{fig:bars}
\end{figure*}

\begin{figure}
  \includegraphics[width=.5\textwidth]{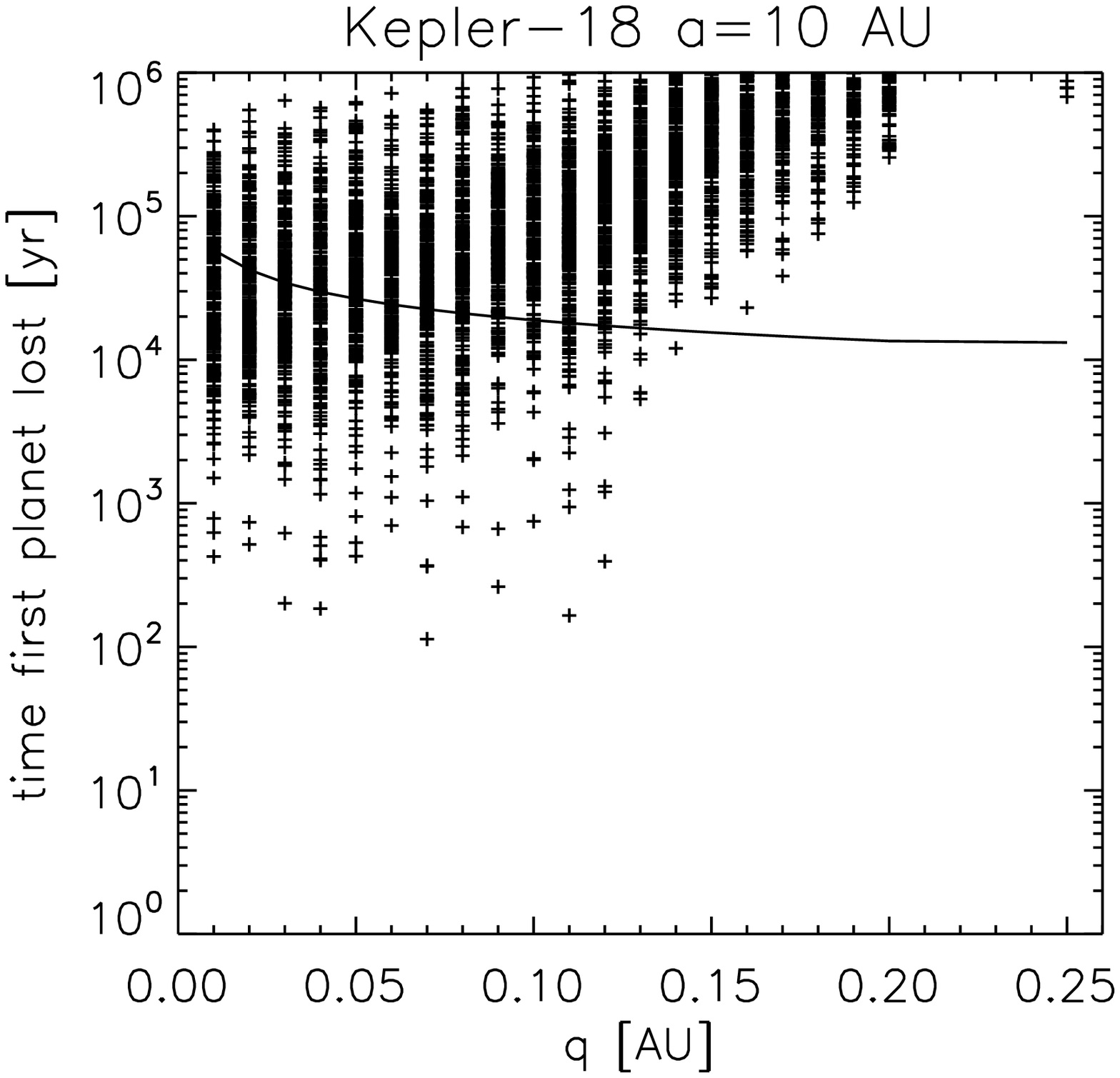}
  \caption{\textbf{Time until loss of the first planet in our Kepler-18 runs with an initial giant planet semimajor axis of 10\,AU.} We overplot the time spent at each pericentre when undergoing Kozai cycles from a $1M_\odot$ perturber at 1000\,AU: in many cases, planets are lost within the time it would take the giant's pericentre to pass through the relevant region.}
  \label{fig:tloss}
\end{figure}

For our chosen systems, we explore varying the giant planet's pericentre and semimajor axis (Fig~\ref{fig:bars}). So long as the giant's orbit is intersecting at least one of the inner planets', the majority of integrations lead to one of the two outcomes in less than 1\,Myr (Fig~\ref{fig:tloss}). For tidal circularisation of the giant's orbit to form a true hot Jupiter, the pericentre must be a few hundredths of an AU (see below); within this distance, nearly all of our simulations result in one of the two outcomes of ejection of the giant or destruction of all the inner planets. The overwhelming outcome is that the three inner planets are destroyed, most commonly by collision with the star, although in some cases the giant accretes one or more, which may significantly enrich the core of the giant. Indeed, many hot Jupiters are observed to have enriched cores \citep[e.g.,][]{Guillot+06}. After evolving our surviving giant planets under tidal forces, we find that giants that have accreted one or more inner planets are more likely to become hot Jupiters than those that have not, although the relative infrequency of collision means that most of the hot Jupiters we form have not accreted other planets (Table~\ref{tab:accrete}).  Due to the extreme collision velocities at these small orbital radii, the giant's radius may be inflated by colliding with a smaller planet \citep{Ketchum+11}, while the impact velocity when two smaller planets collide can be several times their escape velocity, meaning that collisions may generate copious debris \citep{LeinhardtStewart12}. The  inclination of the giant planet with respect to the inner planets does not have a strong effect on the outcome, although with a retrograde giant the fraction of destroyed inner planets colliding with the giant rather than the star rises slightly, as does the number of coexisting systems when the initial $q<0.10$ AU. 

\begin{deluxetable}{lcc}
  \tabletypesize{\scriptsize}
  \tablecaption{Numbers of planets forming hot Jupiters in each system after 10\,Gyr of tidal evolution.\label{tab:accrete}}
  \tablewidth{0pt}
  \tablehead{
    \colhead{ } & \colhead{Form HJ} & \colhead{Do not form HJ}
  }
  \startdata
  \textbf{Kepler-18, $a=10$\,AU} & & \\
  Accreted planet & 146 (22\%) & 509 (78\%) \\
  Did not accrete planet & 150 (13\%) & 987 (87\%)\\
  \hline
  \textbf{Kepler-18, $a=1.25$\,AU} & &\\
  Accreted planet & 204 (36\%) & 365 (64\%) \\
  Did not accrete planet & 407 (11\%) & 3174 (89\%)\\
  \hline
  \textbf{Kepler-23, $a=10$\,AU} & &\\
  Accreted planet & 90 (17\%) & 445 (83\%)\\
  Did not accrete planet & 322 (12\%) & 2344 (88\%)\\
  \hline
  \textbf{Kepler-58, $a=10$\,AU} & &\\
  Accreted planet & 49 (12\%) & 361 (88\%)\\
  Did not accrete planet & 178 (8\%)& 2108 (92\%)\\
  \hline
  \textbf{Kepler-339, $a=10$\,AU} & &\\
  Accreted planet & 138 (26\%) & 384 (74\%)\\
  Did not accrete planet & 434 (22\%) & 1559 (78\%)\\
  \hline
  \textbf{Kepler-339, $a=1.25$\,AU} & &\\
  Accreted planet & 184 (40\%) & 280 (60\%) \\
  Did not accrete planet & 507 (14\%) & 3114 (86\%)
  \enddata
  \tablecomments{``Hot Jupiter'' is here defined to be a planet that becomes tidally circularised to $e<0.1$.}
\end{deluxetable}

\begin{figure}
  \includegraphics[width=.5\textwidth]{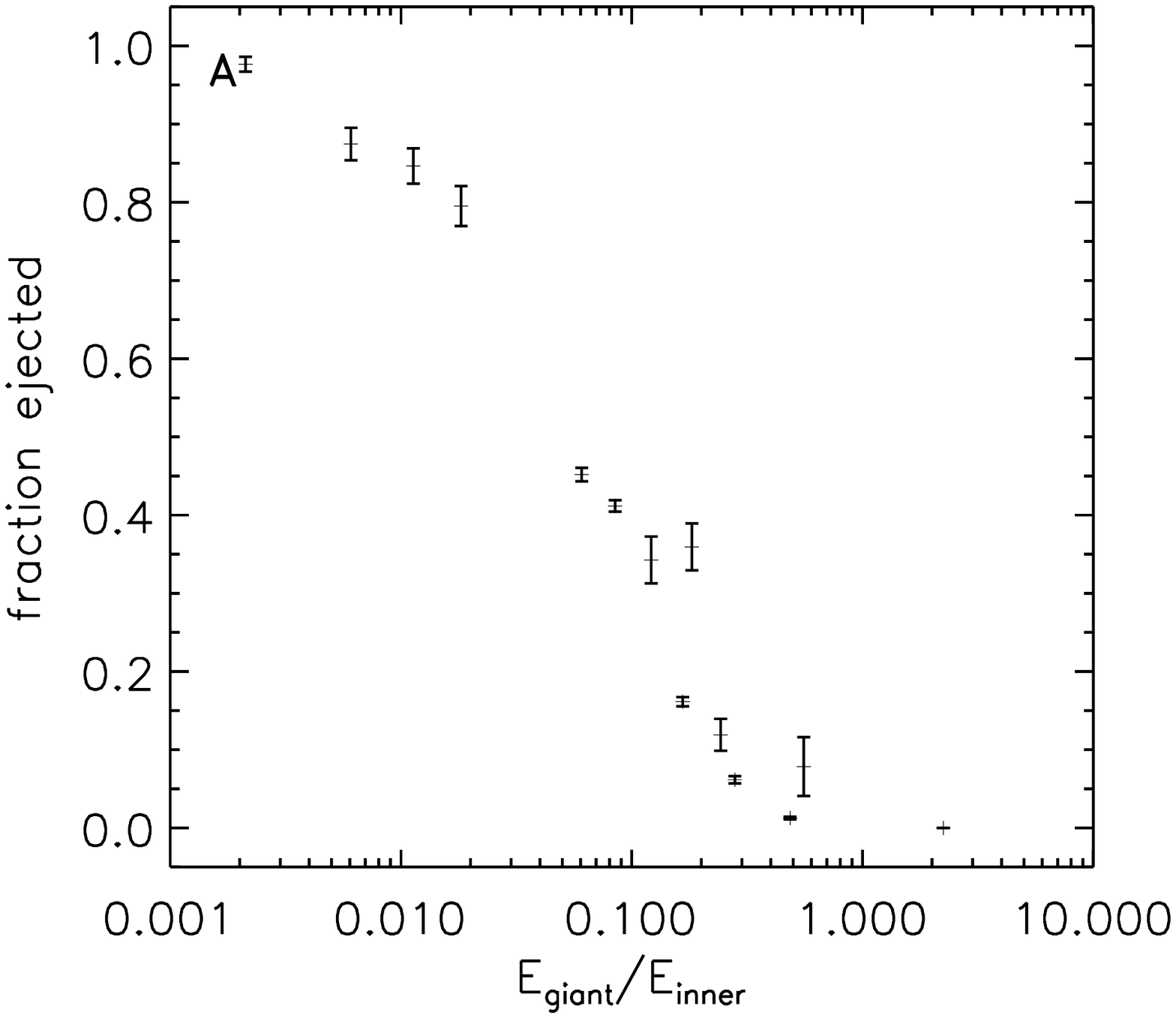}
  \includegraphics[width=.5\textwidth]{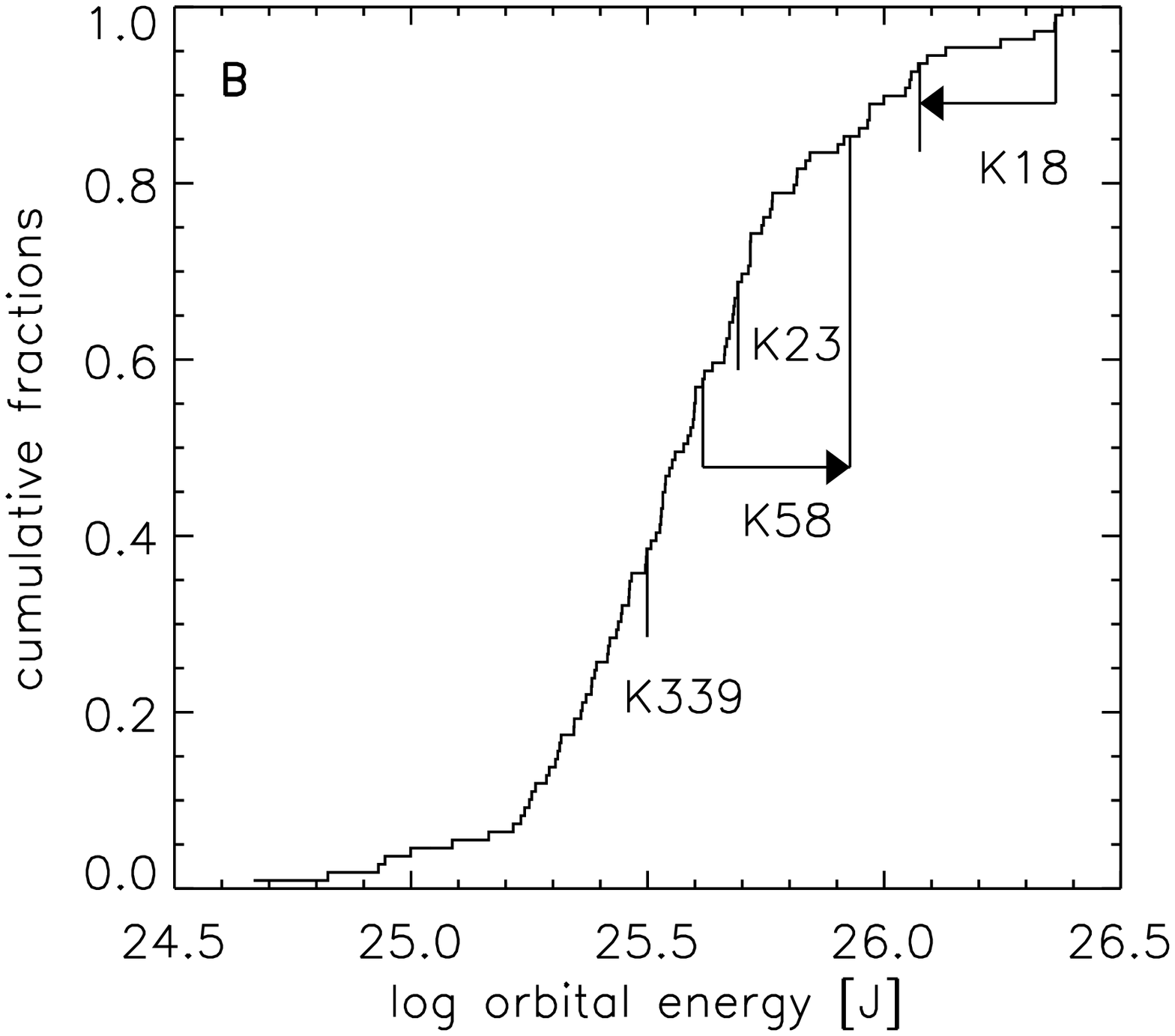}
  \caption{\textbf{Effect of relative orbital energies of the inner planets and the giant on the ejection probability. A:} Chance of ejecting the giant rises as the orbital energy of the the giant planet relative to the inner planets falls. \textbf{B:} Where our exemplar systems sit in the population of \textit{Kepler} triple-transit systems \citep[masses estimated using a mass--radius relation,][]{WM14}. For Kepler-18 and Kepler-58, we mark two values, corresponding to the masses estimated from the mass--radius relation (tail of arrow) and those corresponding to TTV/RV measurements (head of arrow), which were the ones used in the integrations.}
  \label{fig:energy}
\end{figure}

While the ejection of a Jovian planet by Neptune-sized ones may seem surprising, the ratio of ejections of the incoming giant to destruction of the inner planets can be understood in terms of the orbital energies of the two components (Fig~\ref{fig:energy}): as the orbital energy of the giant is decreased (whether through lower mass or through higher semimajor axis), ejection becomes more likely. Planets scattering from near-circular orbits at semimajor axes of $\sim0.1$\,AU would be in a regime favouring collisions, as their physical radius is a significant fraction of their Hill radius \citep{Johansen+12,Petrovich+14}; equivalently, the ratio of their escape velocity to orbital velocity is small, meaning that orbits are not perturbed as much during close encounters. However, for the highly-eccentric planets we consider here, ejection is easily achieved because a small transfer of energy from the inner planets to the giant can lead to a significant change in the latter's semimajor axis. Ejection of the giant is a common outcome for the most massive systems of inner planets we consider when the giant comes in on a wide orbit, but is rare when the inner planets are less massive or the giant's semimajor axis is smaller.

\begin{figure*}
  \vspace{-1cm}
  \includegraphics[width=.86\textwidth]{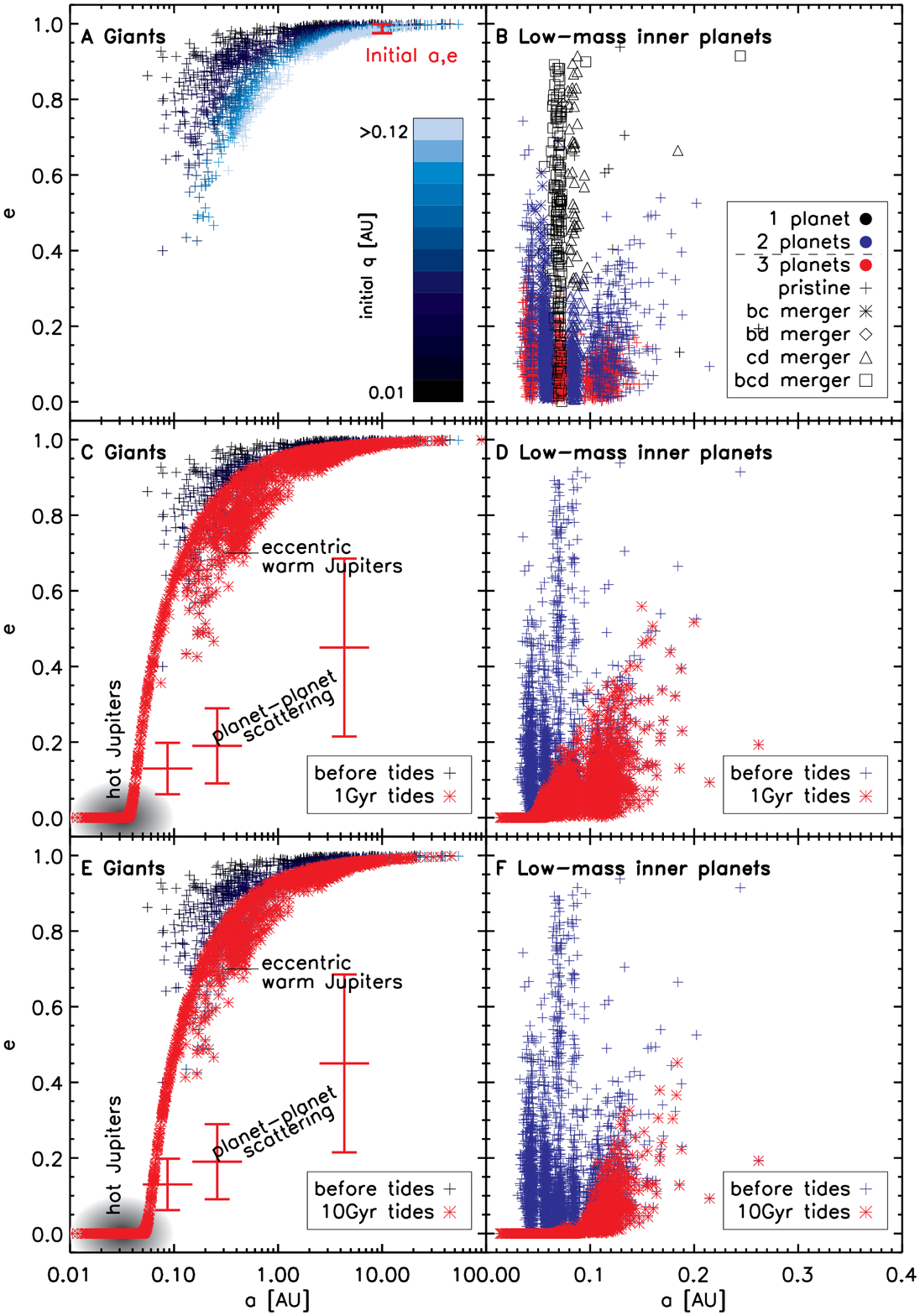}
  \caption{\textbf{Orbital elements at the end of our integrations, for the Kepler-18 system with initial giant planet's semimajor axis of 10 AU. A:} Orbital elements of the giant planet when all inner planets are destroyed. Lighter symbols mark larger initial pericentre $q$. The red line marks the location of the giant planets at the beginning of the integrations. \textbf{B:} Orbital elements of surviving inner planets in systems where the giant was lost, marking multiplicity and merger state of the surviving systems. \textbf{C:} Evolving the distribution of giant planets from panel A under tidal forces for \textbf{1} Gyr. Horizontal red lines mark the mean eccentricity attained through planet--planet scattering \citep{Petrovich+14}, together with associated $1\sigma$ error bars assuming a Rayleigh distribution. We can produce highly-eccentric warm Jupiters at $a<1$\ AU with eccentricities higher than achievable through \emph{in-situ} scattering of giant planets starting at these separations on circular orbits. \textbf{D:} Evolution of the surviving low-mass planets under tidal forces for 1 Gyr. The high eccentricities excited by interaction with the giant are unlikely to persist for the star's main-sequence lifetime. \textbf{E, F:} As C and D but for 10 Gyr of tidal evolution.}
  \label{fig:a-e}
\end{figure*}

Giant planets that destroy the inner planets experience some change to their orbital elements (Fig~\ref{fig:a-e}). Pericentres may change slightly, while semimajor axes may be significantly reduced. Many of the surviving giants maintain the small pericentres needed for tidal circularisation, and will become hot Jupiters after long-term tidal evolution: after 10\,Gyr, between 8\% and 23\% of giants in our integrations circularise to $e<0.1$, depending on the inner planet configuration and the initial semimajor axis of the giant planet (see Table~\ref{tab:accrete}). The final semimajor axes of these hot Jupiters are $\lesssim0.06$\,AU, implying pre-circularisation pericentre distances (after interaction with the inner planets) of $\lesssim0.03$\,AU.

\begin{deluxetable}{lcc}
  \tabletypesize{\scriptsize}
  \tablecaption{Weighted numbers of hot and warm Jupiters from our simulations, after 10Gyr of tidal evolution.\label{tab:hot-warm}}
  \tablewidth{0pt}
  \tablehead{
    \colhead{System} & \colhead{$a<0.1$ AU} & \colhead{$0.1$ AU $<a<0.76$ AU}
  }
  \startdata
  Kepler-18               & 10675 & 4732 \\
  Kepler-18, $a=1.25$ AU  & 23108 & 8474 \\
  Kepler-23               & 14740 & 6184 \\
  Kepler-58               &  8292 & 4024 \\
  Kepler-339              & 20273 & 6020 \\
  Kepler-339, $a=1.25$ AU & 24557 & 7605 \\
  \hline
  KOIs                    &   189 &  125 \\
  RV-detected             &   677 &  255 \\
  RV-detected, $e>0.4$    &       &   80
  \tablecomments{We weight each planet by $1/(a(1-e^2))$ to correct for the geometric transit probability. Below the line we show the numbers of KOIs with $R>8R_\oplus$ from the Q1--Q16 KOI list with the \textit{L+11 \& no FPs} cuts, as well as a sample of RV-detected giant planets (see text).}
\end{deluxetable}

We also form a population of giant planets with large pericentres ($q\gtrsim0.05$\ AU, too large for tidal circularisation), relatively small semimajor axis ($a\lesssim1$\ AU), and moderately high eccentricity ($e\gtrsim0.5$). It is hard to populate this region through \emph{in-situ} scattering of close-in giant planets that may have migrated through a protoplanetary disc \citep{Petrovich+14}, since planets scattering from these semi-major axes are inefficient at exciting eccentricity from circular orbits as they are in a regime favouring collisions; see the discussion above and \cite{Petrovich+14}. Nor is it straightforward to populate this region through tidal circularisation, as the planets lie below the tidal circularisation tracks down which planets move on astrophysically interesting timescales. Planets may enter this region as a result of ``stalled'' Kozai migration, if the parameters of the body driving the planet's eccentricity are sufficient to continue driving Kozai cycles during the tidal dissipation process \citep{Dong+14,DawsonChiang14}, or as a result of secular chaos \citep{WuLithwick11}. Either of these pathways entails certain constraints on the perturber exciting the eccentricity, and in particular it is not clear to what extent the conditions needed to trigger secular chaos are met in practice \citep{Davies+14}. Our model of a high-eccentricity giant interacting with inner planets permits us to populate this same region, without relying on suitable parameters of the exciting body. Unfortunately most \textit{Kepler} candidates do not have measured eccentricities, but still we can compare the numbers of giant planets in semi-major axis bins: from our simulations, after 10\,Gyr of tidal evolution we find around 2--3 times more giant planets in the range $a\in(0,0.1)$ AU than in $a\in(0.1,0.76)$ AU (corresponding to a 240d period) after correcting for the geometrical transit probability\footnote{Ratios are similar if we stop the tidal evolution after 1\ Gyr, although with Kepler-58 and Kepler-18, $a=10$\ AU the ratio is a little lower at 1.7.}; while in our \textit{Kepler} sample, we find only around 50\% more (Table~\ref{tab:hot-warm}). Hence, our results are consistent with the observed population, as we might expect the ``warm Jupiter'' region beyond $0.1$ AU to be populated to some extent by disc migration \citep{Lin+96}---which better explains the low-eccentricity warm Jupiters---while some of the hot Jupiters may be destroyed as a result of tides raised on the star \citep{VR14}. We can also consider the planet population detected by radial-velocity (RV) surveys. A query of the Exoplanet Orbit Database \citep[http://exoplanets.org/,][accessed 2015-05-16]{Han+14} revealed 354 RV-detected planets with masses above $0.3\mathrm{M_J}$. Of these, 33 lie within 0.1\ AU and 59 between 0.1 and 0.76\ AU, 13 of the latter having $e>0.4$. When weighted by their geometric transit probability, this sample has a higher fraction of hot to warm Jupiters than the KOI sample (see Table~\ref{tab:hot-warm}), more in line with the ratio from our simulations. However, if we divide the warm Jupiters into two eccentricity bins at $e=0.4$ (above which \emph{in-situ} scattering is inefficient at exciting eccentricity \citep{Petrovich+14}, and below which tidal circularisation and/or interaction with the inner planets cannot reach), we find over 8 times as many hot Jupiters as eccentric warm Jupiters. This may point to a contribution from disc migration to the low-eccentricity warm Jupiter and hot Jupiter populations, although a detailed treatment of the differences between the RV and KOI samples is beyond the scope of this paper.

\begin{figure*}
  \includegraphics[width=\textwidth]{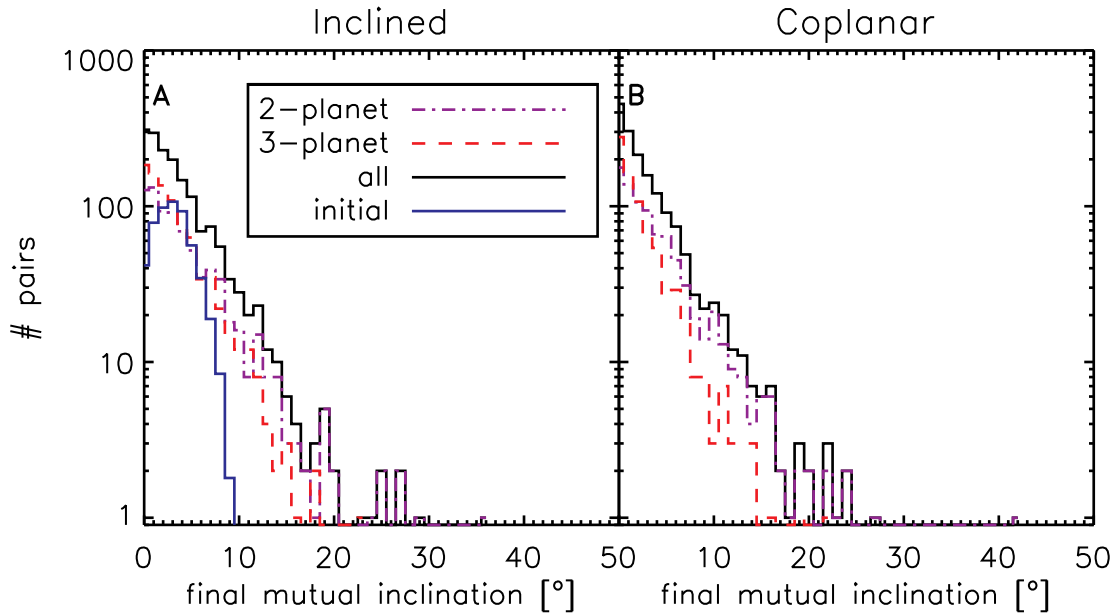}
  \caption{\textbf{Distributions of mutual inclinations of surviving inner planets in Kepler-18 systems that lost their giant. A:} With a spread of mutual inclinations up to $10^\circ$. We also mark the initial distribution (not to scale). \textbf{B:} Initially coplanar inner planets. In both cases the inner planets are not usually significantly excited, although there is a tail of inclinations above $10^\circ$. The distributions arising from the initially coplanar case and the initially inclined case are not significantly different.}
  \label{fig:inc}
\end{figure*}

When the incoming giant is ejected, the inner planets often experience some perturbation. Collisions of inner planets with each other or with the star are common, and the interaction with the giant often leaves systems with only one or two of the original three planets. The eccentricities of inner planets can be strongly excited (Fig~\ref{fig:a-e}), and single survivors in particular can reach very high eccentricities. However, these eccentricities may not survive in the long term, as tidal circularisation acts on the planets' orbits on long timescales: Fig~\ref{fig:a-e} shows that most eccentricities will decay to zero within 10\,Gyr. In contrast, mutual inclinations of the inner planets are not strongly affected, although very flat systems do not retain their coplanarity: initially flat and moderately-inclined (few degrees) systems show similar inclination distributions after ejection of the giant (Fig~\ref{fig:inc}). 

\begin{figure*}
  \includegraphics[width=0.95\textwidth]{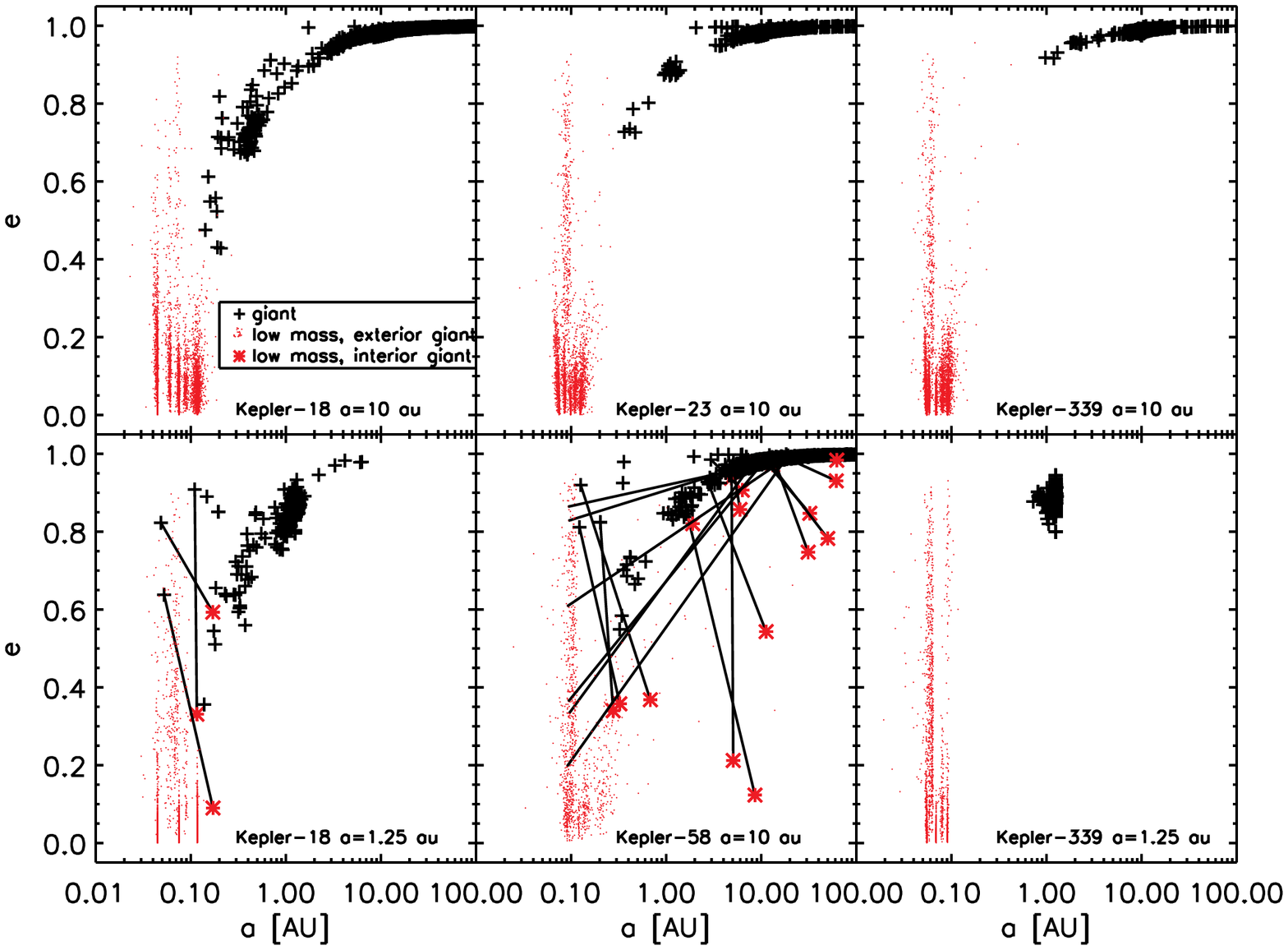}
  \caption{Final (neglecting tidal effects) orbital elements of the giant planets (black) and low-mass inner planets (red) in the integrations in which both the giant and at least one of the inner planets are present after 1\ Myr. ``Inner'' planets that now lie exterior to the giant are marked with large red stars, and planets in these systems are linked with black lines.}
  \label{fig:coexist}
\end{figure*}

Finally, we discuss the systems where the giant and at least one low-mass inner planet coexist at the end of the integration, shown in grey in Figure~\ref{fig:bars}. In the overwhelming majority of these systems, the giant planet's pericentre lies beyond the orbit of the outermost inner planet, and the final coexisting system looks much like the initial setup, retaining a highly-eccentric giant on a wide orbit with one or more low-mass planets close to the star; the number of inner planets may however be depleted by collisions. Interestingly, we do find a very few cases (23 in number) where at the end of the integration the giant planet's semi-major axis is smaller than that of the outermost surviving inner planet. In 17 of these cases, all from the Kepler-58 simulations, both the giant and planet~d lie beyond 1\ AU; in 6 of these, an additional planet remained at $\sim0.09$\ AU. In the remaining 6 cases, three each in the Kepler-58 simulations and the Kepler-18 simulations with the giant starting at 1.25\ AU, the giant planet has collided with a b--c merger product, and the specific energy of the resulting body is sufficiently low that its orbit lies interior to that of planet~d. Five of these systems survived integration for 10\ Myr, and in all cases the giant planet's pericentre is sufficiently small as to permit tidal circularisation and the formation of a hot Jupiter. In these five systems, the mutual inclination is very high, oscillating around $90^\circ$ and hampering the prospects for detection of both planets by transit photometry. However, these hot Jupiters with surviving companions form only 0.5\% of hot Jupiters formed in the Kepler-18, $a=1.25$\ AU integrations and 0.9\% of those formed in the Kepler-58 integrations. Hence, while survival of companion planets is possible given the right conditions (viz.\ sufficiently massive inner planets), it occurs in only a tiny fraction of even these systems. We show the semi-major axes and eccentricities of planets in these coexisting systems in Figure~\ref{fig:coexist}, highlighting the few systems in which the giant planet lies interior to one of the smaller ones.

\newpage

\section{Discussion}

\subsection{The Evolutionary Context of our Simulations}

Our method assumes that the evolution of the system can be broken into three stages: an initial stage of excitation of the giant planet's eccentricity; interaction of the highly-eccentric giant with any inner planets; and subsequent tidal circularisation of surviving planets' orbits. We do not explicitly treat the initial phase, since the parameter space of mechanisms and perturbers is very large, and the combination of the small integrator step size needed to resolve the inner planets' orbits conflicts with the long timescales \citep[which can be over $10^8$ years,][]{WuLithwick11} needed to excite the eccentricity of the giant planet. The set-up for our simulations is probably most accurate for a scattering scenario, where the giant planet's pericentre is impulsively changed to a very low value. In a Kozai or other secular scenario with a smoother eccentricity excitation, it is likely that the secular evolution will either continue until the timescale for secular evolution is comparable to the timescale for interaction between the inner planets and the giant (see Fig~\ref{fig:tloss}), at which point our integrations begin; or that the interaction with the inner planets briefly halts the secular cycle until they are destroyed \citep[similar to the effects of general relativistic precession,][]{WuMurray03}, after which the secular cycle may resume. Note that destruction of the inner planets can occur at pericentres wider than those at which the giant's orbit actually overlaps the inner planets' (Fig~\ref{fig:bars}).

We also neglect any further effect of the body perturbing the giant planet during our integrations and after they have finished. This is again most accurate for a scattering scenario, where the swift reduction of the giant's apocentre following interaction with the inner planets would decouple the giant planet from its original perturber. In a secular or Kozai scenario, the secular cycles may resume after the inner system has been cleared, which will affect the statistics of hot and warm Jupiters we have estimated (Tables~\ref{tab:accrete} and~\ref{tab:hot-warm}). Following the entire evolution of these systems from initial eccentricity forcing through to final tidal circularisation would be a fruitful avenue of future research.

Although we have not treated the full evolution of these systems in this work, we can attempt to relate the outcomes of the integrations to the eccentricity excitation mechanism. In particular, a large semi-major axis of the giant planet increases the probability that it will be ejected instead of destroying the inner system. Driving a planet's pericentre to very small distances by scattering from very wide orbits (note that to achieve a semi-major axis of 10\ AU, the scattering event would have to take place at around 20\ AU) is difficult \citep{Mustill+14}, and the giant planets that we find vulnerable to ejection when they interact with the inner planets may be more likely to have been excited by Kozai perturbations from a wide binary companion.

\subsection{Robustness of our Findings}

The main result of our study---that giant planets with sufficient orbital eccentricity to become hot Jupiters destroy low-mass inner planets in the system---is robust to the masses of these inner planets, so long as they are not so massive as to eject the giant. In the absence of damping mechanisms that can separate and circularise orbits, the intersecting orbits of the giant and the inner planets lead to either collisions or ejections until orbits no longer intersect. In contrast, in very young systems, eccentricity can readily be damped by the protoplanetary gas disc or by massive populations of planetesimals, helping to explain why Type~II migration of giant planets does not totally suppress the formation of other planets in the inner parts of these systems: bodies thrown out by the giant can recircularise and accrete outside its orbit \citep{Mandell+07}.

In our systems, in contrast to systems during the protoplanetary disc phase, gas is no longer present, and massive planetesimal populations are impossible to sustain close to the star for long time-scales \citep{Wyatt+07}. Two additional sources of damping may play a role in these systems. First, debris may be generated in hypervelocity collisions between inner rocky planets, but integrations with the mass of the Kepler-339 planets distributed among 100 smaller bodies did not show significant damping of the giant's eccentricity. Second, tidal circularisation acts, but on timescales much longer than the time for planet--planet interactions to end in our systems.

\section{Conclusions}

We have shown that high-eccentricity migration of a giant planet to form a hot Jupiter necessarily leads to the removal of any pre-existing planets on orbits of a few tenths of an AU in the system, thus accounting for the observed lack of close companions to hot Jupiters. This supports a high-eccentricity migration scenario for hot Jupiters, as migration through a protoplanetary gas disc usually does not fully suppress planet formation \citep{FoggNelson07a,FoggNelson07b,Mandell+07,Ketchum+11,Ogihara+14}. We find that under high-eccentricity migration, when the giant's pericentre is sufficiently small to permit tidal circularisation, either the giant or the inner planets must be lost from the system. A very small fraction ($<1$\% even with favourable parameters) of the hot Jupiters we form do end up interior to a surviving low-mass planet, but this outcome is very uncommon: if such a low-mass close companion to a hot Jupiter were in future to be found, it would mean that in that system at least the migration almost certainly proceeded through a disc. When the giant planet does destroy the inner system, the interaction sometimes raises the pericentre of the eccentric giant planets sufficiently to prevent tidal circularisation, providing a novel way of producing eccentric warm Jupiters; other giants whose pericentres are initially too high for tidal circularisation may be brought to populate the same region as they lose energy due to interaction with the inner planets.

It is unknown which mechanism of eccentricity excitation dominates, be it scattering, the Kozai effect, or low-inclination secular interactions, but we expect that the inability of inner planets to survive in systems forming hot Jupiters will remain a robust result when future simulations coupling the evolution of the outer system, driving the giant's eccentricity excitation, and the inner system are performed.

\newpage

\acknowledgments

We thank Sean Raymond, Cristobal Petrovich, and the anonymous reviewer for comments on the manuscript. This work has been funded by grant number KAW 2012.0150 from the Knut and Alice Wallenberg foundation, the Swedish Research Council (grants 2010-3710 and 2011-3991), and the European Research Council starting grant 278675-PEBBLE2PLANET. This work has made use of computing facilities at the Universidad Aut\'onoma de Madrid.

\bibliography{3pplusj}
\bibliographystyle{apj}

\end{document}